\documentclass[twocolumn,showpacs,showkeys,preprintnumbers,amsmath,amssymb,aps,prb]{revtex4-1}

\usepackage{graphicx}% Include figure files
\usepackage{dcolumn}% Align table columns on decimal point
\usepackage{bm}% bold math
\begin{document}

%\preprint{LA-UR-13-21231 (version 2)}

\title{Thermal electronic excitations in liquid metals}

\author{Eric D. Chisolm}
\author{Nicolas Bock}
\author{Sven P. Rudin}
\author{Duane C. Wallace}

\affiliation{Theoretical Division, Los Alamos National Laboratory, Los Alamos, New Mexico 87545}

\date{\today}

\begin{abstract}
Thermal electronic excitations in metal crystals are calculated by starting with a reference structure for 
the nuclei:  the crystal structure of the appropriate phase.  Here we explain the corresponding theory for
metal liquids, starting with an appropriate reference structure for a liquid.  We explain the significance
of these structures, and we briefly review how to find them and calculate their properties.  Then we examine 
the electronic densities of states for liquid structures of Na, Al, and Cu, comparing them to their crystal 
forms.  Next we explain how to calculate the dominant electronic thermal excitation term, considering issues
of accuracy that do not arise in the crystal theory.  Finally we briefly discuss the contribution from the interaction
between excited electrons and moving nuclei.
\end{abstract}


\maketitle

\section{Introduction}

For metallic crystals, thermal electronic excitation theory is well developed.  The leading 
approximation is to fix the nuclei at a reference structure, namely the appropriate crystal structure, 
at which we calculate the electronic density of states (DOS) and use statistical mechanics to 
describe the excitation of electrons across the Fermi level.  It then remains to describe the 
correction due to the vibrational motion of the nuclei.  This correction is expressed through 
electron-phonon interactions, and is treated in perturbation theory to the first order in which 
its free energy contribution is nonzero.\cite{M_SPJ58a,E_SPJ60a,G_PS76a,A_PRB87a,BCW_PRB05a,
BWC_PRB06a}  The total crystal free energy then consists of a series 
of diminishing contributions: the structural potential, nuclear vibrations, electronic excitation, 
and electron-phonon interactions, where the last contribution is the only truly complicated one, and is 
negligible for practical equation-of-state applications.  The same ordering of terms holds for 
nearly-free-electron (NFE) metals and transition metals, though the electronic excitation effects are 
much larger in transition metals.\cite{EWW_PRB92a}

A similar formulation has not been developed for liquid metals because the first step requires a 
reference structure, and such a structure has not been available for liquids.  However, the needed 
structure naturally appears in the V-T (vibration-transit) theory of liquid dynamics.\cite{CW_JPCM01a,W_02a} 
In this theory, the potential energy valleys which underlie the liquid 
motion are divided into symmetric and random classes, where the numerically superior random valleys 
dominate the liquid motion, and all have the same statistical mechanical properties in the thermodynamic 
limit.  The nuclear motion consists of vibrations in a random valley interspersed with transits which 
carry the system between random valleys.  A transit is accomplished by a small local group of nuclei, 
and transits are going on at a high rate throughout the liquid.  The nuclear configuration at any instant 
is within a single random valley, so the valley minimum, called a random structure or liquid structure, 
serves as the reference structure for the nuclear motion.

Our goal is to explain how to use the liquid structure electronic DOS to study the statistical mechanics 
of thermal electronic excitations in liquids.  We start with density functional theory (DFT) results for the liquid 
structure DOS for Na, Al, and Cu, and we show the close similarity between liquid and crystal DOS.  Then 
we discuss the electronic excitation theory, which is formally the same for crystal and liquid with an important 
exception: the liquid's higher temperature range requires the integral formulation.  With the theory 
in hand, we then estimate the size of electronic excitation effects: they are small for the crystal at all temperatures, 
and are nearly the same for liquid and crystal at melt, but they can become important in the 
high-temperature liquid.  Finally, we observe that the interaction between electronic excitations and nuclear 
motion will also become non-negligible in the liquid at high temperatures.

\section{The liquid structure electronic DOS}

Since the use of the liquid structure DOS is new in liquid dynamics theory, we will describe 
the procedure for finding these structures and explain how we have validated their application to V-T theory.  
We also distinguish two classes of melting of elements: normal melting, where liquid and crystal 
have essentially the same electronic structure, and anomalous melting, where the electronic structures
are significantly different.\cite{W_PRSLA91a}  Anomalous melting is rare, and it breaks the rule we 
commonly apply to metallic systems, that liquid and crystal are similar at melt.  For simplicity of discussion 
here, we shall consider only normal melting metallic systems.

Finding liquid structures proceeds in two steps.  First, random structures are found by quenching an $N$-atom 
supercell to a potential energy minimum, typically through direct minimization of the total energy (using e.g.\ 
a DFT code such as VASP\cite{KF_PRB96a,KJ_PRB99a}).  Because of the dominance and uniformity 
of the random valleys, quenches almost always arrive at random structures, and these structures possess 
very narrow distributions\cite{DW_PRE07a} of the Hamiltonian parameters: the structural potential $\Phi_0$ and 
the vibrational characteristic temperatures $\theta_n$ for $n = 0, 1, 2$.  In our experience, a set of $10$ 
or more DFT quenches from independent stochastic configurations\cite{HBPLDCW_PRE09a,HBPCLDW_PRB10a} 
will have a distribution of $\Phi_0$ lying above the crystal potential by roughly $k_BT_m$, and of width 
$\lesssim 0.04\,k_BT_m$ for $N \gtrsim 150$.  The $\theta_n$ are likewise narrowly distributed.  These 
properties will verify the structures are random.  Second, the electronic properties associated with the random 
structure must be checked.  For systems considered here, the liquid is metallic, so the electronic structure 
should be the correct metallic structure.  Random structures having the proper liquid electronic structure are 
called liquid structures.

Because the nuclear motion in elemental liquids poses an intricate mechanical problem, V-T theory has been 
developed primarily through comparison of statistical mechanical models with experimental data.  This 
development has provided a level of verification of the role of the liquid structure in liquid dynamics theory.
\cite{CW_JPCM01a}  The most convincing validation of the liquid structure is now provided by DFT 
calculations which, together with a statistical mechanical model for the small contribution from transits,
\cite{WCBD_PRE10a} yield highly accurate thermodynamic properties of the liquid at melt for Na and Cu.
\cite{BHPLCDW_PRB10a}

Finally, there are independent theoretical arguments that apply to the liquid DOS.  In pseudopotential 
perturbation theory, the electronic DOS is free electron in zeroth order, and the presence of the ions makes 
a small correction;\cite{H_66a} this theory applies to Na and Al.  Moreover, for normal melting elements, which 
include Na, Al, and Cu, the electronic structure is the same for liquid and crystal at the same volume, so the 
DOS are very similar.  Small differences arise because variations in the crystal DOS are smoothed out of the 
liquid DOS by its disordered nuclear arrangement (see Figures below).

The DFT calculations employ VASP, use the PW91 generalized gradient approximation\cite{PCVJPSF_PRB92a} (GGA),
and treat the valence electrons in a plane-wave basis with element-specific projector-augmented wave potentials.
\cite{B_PRB94a}  The calculations employ first order Methfessel-Paxton smearing\cite{MP_PRB89a} with width $0.2$ eV
and an energy cutoff of $102$ eV (Na), $301$ eV (Al), and $342$ eV (Cu).  The self-consistent field
cycles are converged to within $10^{-6}$ eV.  The crystal DOS in the figures were calculated with a $40 \times 
40 \times 40$ $k$-point mesh.  The liquid DOS for Na and Al were calculated with a $12 \times 12 \times 12$ 
mesh, while a $7 \times 7 \times 7$ mesh was used for Cu.  For Na, $N=500$, while $N=150$ for Al and Cu.

\begin{figure}[h]
\includegraphics[height=.3\textheight,clip=true]{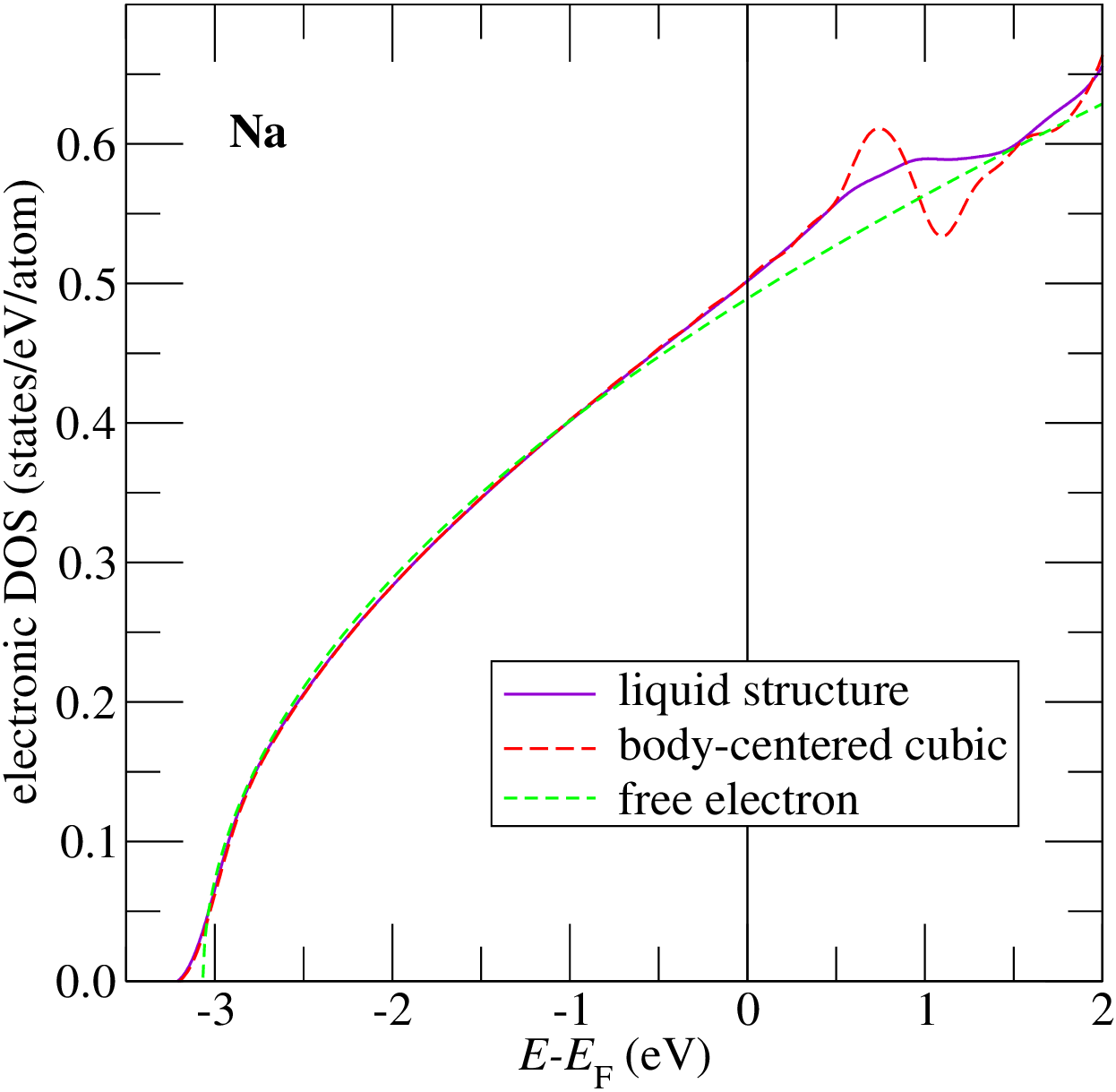}
\caption{(Color online) The electronic density of states for crystal and liquid Na at an atomic volume of $41$ 
\AA$^3$.  (For comparison, the volume of the crystal and liquid at melt are, respectively, $39.79$ and $40.93$ 
\AA$^3$.)  The solid shows Brillouin zone band gaps at $\epsilon > \epsilon_F$, and the liquid shows a smoothed
version of the same feature.  The free-electron DOS with one electron per atom is also shown.}
\label{NaDOS}
\end{figure}

Figure~\ref{NaDOS} shows that the calculated electronic DOS for Na agrees with the free electron 
result to high accuracy for the crystal at $\epsilon \lesssim \epsilon_F$, and for the liquid at all energies, as 
expected.  Effects from Brillouin zone band gaps appear in the crystal at $\epsilon > \epsilon_F$, and similar 
but smoothed effects appear in the liquid.

\begin{figure}[h]
\includegraphics[height=.3\textheight,clip=true]{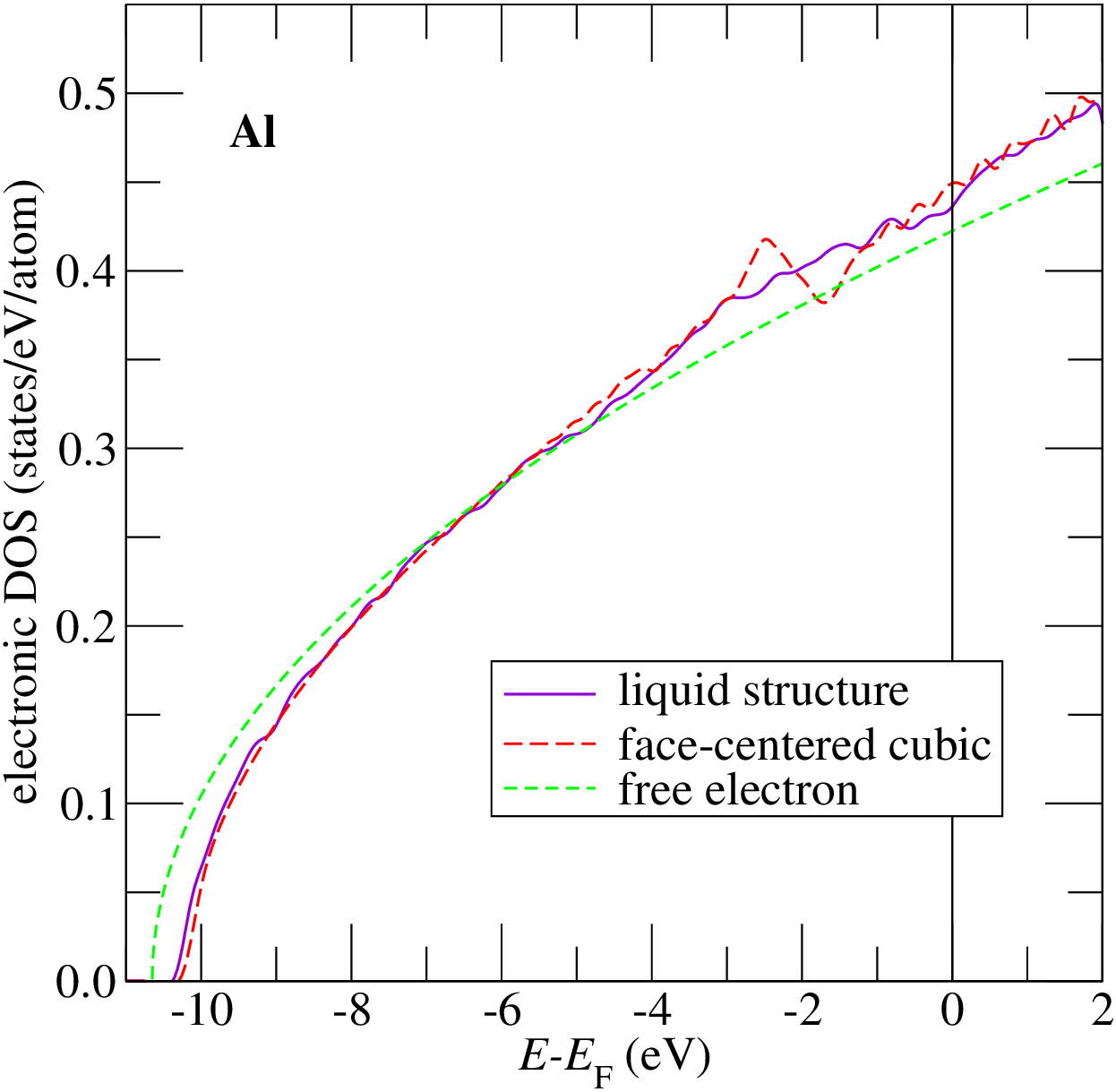}
\caption{(Color online) The electronic density of states for crystal and liquid Al at an atomic volume of $19$ 
\AA$^3$.  (For comparison, the volume of the crystal and liquid at melt are, respectively, $17.56$ and $18.78$ 
\AA$^3$.)  The solid (but not the liquid) shows Brillouin zone band gaps at $\epsilon < \epsilon_F$.  The 
free-electron DOS with three electrons per atom is also shown.}
\label{AlDOS}
\end{figure}

In Figure~\ref{AlDOS} for Al, the free electron DOS is in good overall agreement with DFT for crystal and liquid alike.  
More precisely, the free electron DOS is slightly wider than DFT, and averages around $5\%$ lower at 
energies within $2$ eV of $\epsilon_F$.  Brillouin zone band gaps appear below $\epsilon_F$ in the 
crystal, but are absent in the liquid.

\begin{figure}[h]
\includegraphics[height=.3\textheight,clip=true]{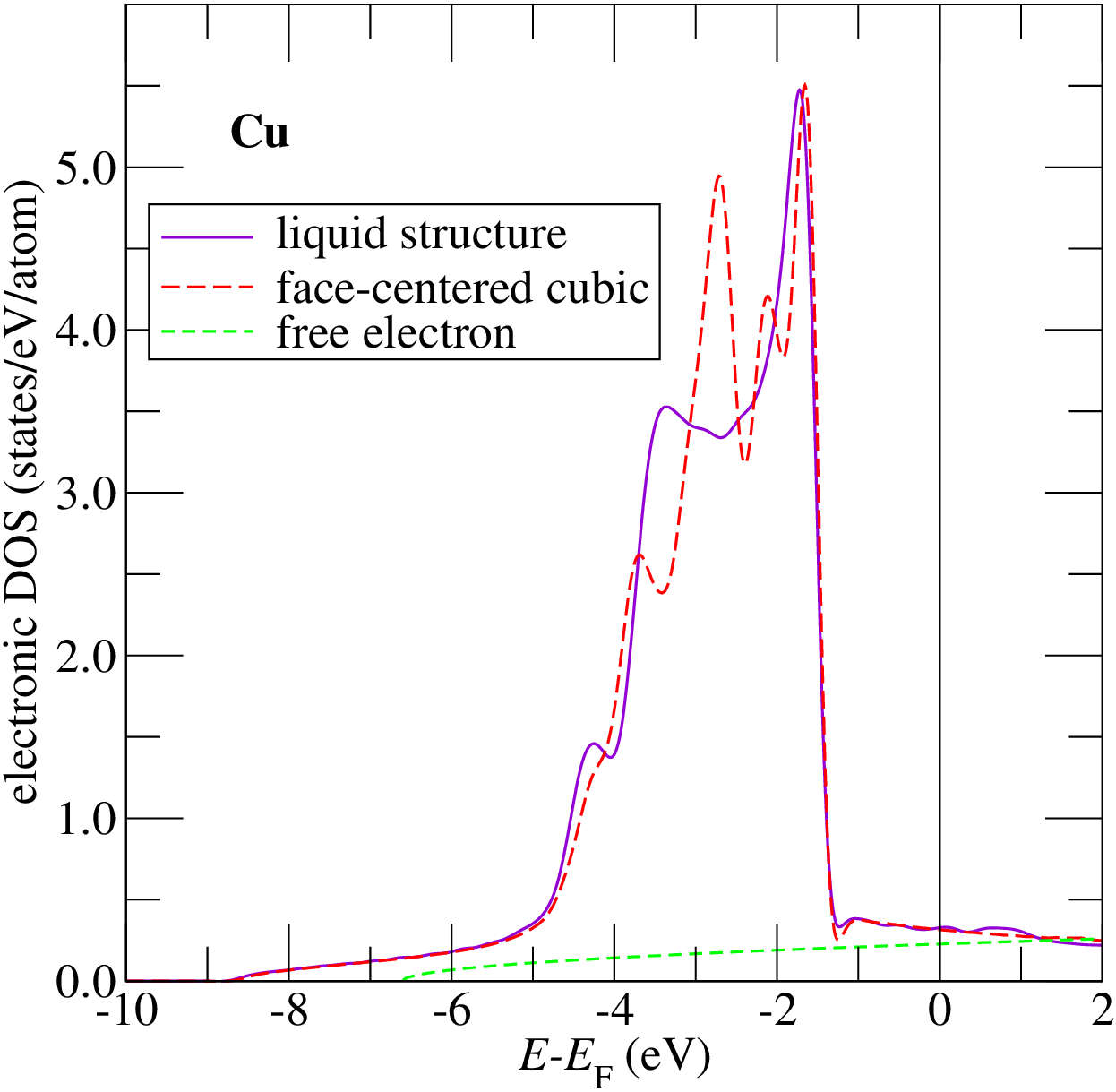}
\caption{(Color online) The electronic density of states for crystal and liquid Cu at an atomic volume of $13$ 
\AA$^3$.  (For comparison, the volume of the crystal and liquid at melt are, respectively, $13.05$ and $13.54$ 
\AA$^3$.)  The bonding and antibonding states separated by a pseudogap are clearly visible, although the 
gap is smoothed in the liquid.  The free-electron DOS with one electron per atom is also shown, agreeing 
approximately at $\epsilon = \epsilon_F$.}
\label{CuDOS}
\end{figure}

In Figure~\ref{CuDOS} for Cu, the crystal shows the well-known $d$-band structure with bonding and antibonding 
states separated by a pseudogap.  The liquid shows the same form with the same width, but noticeably 
smoothed.  We have also used the free electron approximation for electronic excitation effects in the liquid 
noble metals,\cite{BHPLCDW_PRB10a,W_PRSLA91b} a notion that goes back to the Mott and Jones model of 
a filled $d$ band overlapped by a broad partially filled $s$ band (Figure~80 of Ref.~\onlinecite{MJ_58a}).  The free 
electron DOS in Figure~\ref{CuDOS} represents the Mott and Jones $s$ band, which agrees with the crystal 
and liquid DOS to within $30\%$ at the Fermi energy.  This shows that the free electron approximation is adequate 
when the temperature is not too high and the thermal electronic contribution is not needed to very high accuracy.

\section{Statistical mechanics}

To examine the statistical mechanical role of the liquid DOS, we shall discuss the thermodynamic internal 
energy $U(V,T)$.  The estimates we find for the energy hold approximately for the entropy and free energy 
as well.  In V-T theory the internal energy is
\begin{equation}
U = \Phi_0 + U_\textrm{vib} + U_\textrm{tr} + U_\textrm{el} + \delta U_\textrm{el}
\label{Udef}
\end{equation}
where $\Phi_0(V)$ is the structural potential energy introduced earlier, $U_\textrm{vib}(V,T)$ and $U_\textrm{tr}(V,T)$ 
express respectively the nuclear vibrational and transit motions, and $U_\textrm{el}(V,T)$ and 
$\delta U_\textrm{el}(V,T) $ express respectively the electronic excitation with nuclei fixed at the structure, 
and the correction to this due to the nuclear motion.  The zero of energy appears in $\Phi_0(V)$, while the 
remaining thermal excitation contributions are each measured from $\Phi_0(V)$.  This decomposition is 
analogous to the standard decomposition for crystals, with the additional term $U_\textrm{tr}$ representing an
additional type of nuclear motion.

For a monatomic liquid at $T_m$, by far the largest thermal energy contribution is $U_\textrm{vib}$.  
$U_\textrm{tr}$ makes a significant contribution around $0.1\,U_\textrm{vib}$.  $U_\textrm{el}$ is around 
$0.01\,U_\textrm{vib}$ for NFE metals, around $0.1\,U_\textrm{vib}$ for transition metals, and is easily 
evaluated from the liquid DOS, Figures~\ref{NaDOS}-\ref{CuDOS}.  The smallest and most difficult term 
to evaluate is $\delta U_\textrm{el}$.  We shall therefore concentrate first on $U_\textrm{el}$ and consider 
$\delta U_\textrm{el}$ for the high-temperature liquid in the last section.  

For liquid theory, the Sommerfeld expansion is not efficient due to rapid fluctuations in the DOS around the 
Fermi energy, so we use the DOS integral formulation for thermodynamic functions.  In this formulation,
\begin{equation}
U_\textrm{el} = \int_{-\infty}^{\infty} n(\epsilon) \left[ \bar{f}(\epsilon) - g(\epsilon) \right] \epsilon \, d\epsilon
\label{intU}
\end{equation}
where 
\begin{equation}
\bar{f}(\epsilon) = \frac{1}{e^{\beta(\epsilon - \mu)} + 1}.
\label{fbar}
\end{equation}
Here $\mu$ is the chemical potential and $g(\epsilon)$ is the ground state evaluation of $\bar{f}(\epsilon)$.  
The term in $g(\epsilon)$ subtracts the electronic ground state energy, which has already been included in 
$\Phi_0$ (Section 3 of Ref.~\onlinecite{W_02a}).  Volume dependence is contained in $n(\epsilon)$ and $\mu$, 
and we shall estimate temperature effects at constant volume.  This is the formulation used by 
Ref.~\onlinecite{EWW_PRB92a} to calculate the electronic excitation entropy for transition metal crystals to $T_m$.

%Transits have the effect of reducing the vibrational character of the nuclear motion and increasing its diffusive 
%character.  With increasing temperature, this effect becomes more pronounced until a temperature is reached 
%where the nuclear motion cannot be considered vibrational in leading approximation.  Our estimate for this 
%temperature is $4\, \theta_\textrm{tr}$, where $\theta_\textrm{tr}(V)$ is the material-specific scaling temperature 
%for transit effects \cite{} and is listed in Table~\ref{matprops}.  To the best of our knowledge, V-T theory will provide 
%acceptable estimates of liquid properties for $T \lesssim 4\,\theta_\textrm{tr}$.

\section{The liquid to high temperatures}

At a fixed volume $V$, the liquid persists as the stable phase from $T_m(V)$ to an upper limit which we estimate 
as roughly $4\,T_m(V)$.  As temperature increases further, the system enters the broad liquid-to-gas transition 
where neither liquid theory nor gas theory is accurate.\cite{WHJS_PRA82a}  We believe that current V-T theory 
will be a decent approximation to $T \approx 4\,T_m$, but that it will require modification to remain highly accurate 
to such a temperature.

Necessity of the integral formulation, Eq.~\eqref{intU}, is shown by estimating the range of $\epsilon - \mu$ 
that contributes to the integral.  In the approximation that $n(\epsilon)$ is constant, calculation of $U_\textrm{el}$ 
to an accuracy of $2\%$ requires $\beta|\epsilon - \mu|$ from $0$ to $6$.  The corresponding range of 
$|\epsilon - \mu|$ at $T=4\,T_m$ is listed in Table~\ref{matprops}.  We conclude that a significant part of the 
valence electron DOS contributes to $U_\textrm{el}$ for liquid metals at high temperatures.

\begin{table}
\begin{ruledtabular}
\begin{tabular}{l|ccc}
Quantity & Na & Al & Cu \\
\hline
$T_m$ (K)                                                           & $371.0$ & $933.5$ & $1358$ \\
%$\theta_\textrm{tr}$ (K)                                   & $570$    & $980$    & $1358$ \\
$|\epsilon - \mu|$ range at $4 \, T_m$ (eV)  & $0-0.8$  &  $0-2.0$ & $0-2.8$ \\
\end {tabular}
\end{ruledtabular}
\caption{The melt temperature and range of $|\epsilon - \mu|$ needed for accurate evaluation of $U_\textrm{el}$ 
for each material in this study.}
\label{matprops}
\end{table}

We have listed in Table~\ref{estimates} our estimates of the ratio $U_\textrm{el} / U_\textrm{expt}$ to high temperatures, 
starting from the crystal and liquid data tables of Ref.~\onlinecite{W_02a}, for the class of liquid NFE elements and the class of liquid 
transition metals.  They show us that for NFE metals $U_\textrm{el}$ is small enough that one can use the free electron 
DOS for reliable estimates, but for transition metals the liquid DOS is required. At the high temperature of $4\,T_m$, 
$U_\textrm{el} / U_\textrm{expt}$ for NFE metals approaches the level exhibited by transition metals at $T_m$, while 
$U_\textrm{el}$ for transition metals can become a major part of $U_\textrm{expt}$.

%For most elemental metals, the electronic structures of crystal and liquid are similar at a common volume, 
%hence the internuclear forces and electronic DOS are also similar.  For such elements, at $T \approx T_m$
%electronic excitation properties of the liquid can be reliably estimated from crystal data.  We have made such 
%estimates (Tables 19.1 and 19.2 of Ref.~\onlinecite{}) and have used V-T theory to extend them to high temperatures, 
%for the class of NFE metals and the class of transition metals.  The estimates are given in Table~\ref{estimates} 
%and from them we draw these conclusions.

\begin{table}
\begin{ruledtabular}
\begin{tabular}{c|cc}
Quantity & NFE metals & Transition metals \\
\hline
$U_\textrm{el} / U_\textrm{expt}$ at $T_m$          & $0.014 \pm 0.005$ & $0.13 \pm 0.02$ \\
$U_\textrm{el} / U_\textrm{expt}$ at $4\,T_m$      & $0.07$                      & $ \lesssim 0.40$         \\
\end{tabular}
\end{ruledtabular}
\caption{The relative contributions of the thermal electronic excitation energy to the total energy for NFE and 
transition metals at two temperatures.  (The high-temperature value for transition metals depends on where the 
Fermi energy lies with respect to the $d$ band; we give an upper bound.)}
\label{estimates}
\end{table}

%These results summarize a thermal electronic excitation theory for liquids in leading approximation.  
%We find a reference structure, namely the random structure, at which we calculate the electronic DOS and 
%use Eq.~\eqref{intU} (or analogous formulas for other properties) to describe electronic excitations.  The 
%correction due to the vibrational motion of the nuclei, $\delta U_\textrm{el}$, is complicated but negligible 
%for practical equation-of-state applications, as with crystals.  The general trends for NFE and transition metals
%found for crystals also hold for liquids.

\section{Interaction between electronic excitation and nuclear motion}

The term $\delta U_\textrm{el}$ in the internal energy, Eq.~\eqref{Udef}, expresses the interaction between 
electronic excitation and nuclear motion.  In liquid and crystal alike, the formal theory for this term has a nonadiabatic 
contribution dominant at low temperatures and an adiabatic contribution dominant at high temperatures.  
Practically, the adiabatic term dominates in the liquid at $T \geq T_m$.  We have 
$\delta U_\textrm{el} \propto TU_\textrm{el}$ in the crystal, but the $T$-dependence should weaken in the 
liquid.  Our estimate is that the magnitude of $\delta U_\textrm{el}$ will surpass that of $U_\textrm{el}$ in 
the range $T_m \leq T \leq 4\,T_m$.  Hence in contrast with crystal theory, both $\delta U_\textrm{el}$ and 
$U_\textrm{el}$ will become at least significant, and in some cases important, in the high temperature liquid.

In the absence of a liquid reference structure for $n(\epsilon)$, Hafner, Kresse, and coworkers 
\cite{JH_JPCM90a,HJ_PRB90a,HJ_PRB92a,KH_PRB93a,KH_PRB94a,CSHK_PRB03a} presented the 
average DOS $\langle n(\epsilon) \rangle_\textrm{MD}$, averaged over nuclear configurations taken from an 
\textit{ab initio} molecular dynamics (MD) trajectory.  This average DOS is temperature-dependent and contains 
nuclear motion effects, hence is quite different from the liquid structure DOS presented here.  Indeed, the 
two formulations are to a certain extent complementary because $\langle n(\epsilon) \rangle_\textrm{MD}$ 
contains information on $\delta U_\textrm{el}$, the last term in Eq.~\eqref{Udef}.  We shall examine the 
interrelation of the two formulations and shall present a detailed study of the theory for $U_\textrm{el}$ and 
$\delta U_\textrm{el}$ in a forthcoming paper.

\acknowledgments{We appreciate helpful comments from Erik Holmstr\"{o}m and John Wills. This work was 
supported by the U.\ S. DOE under Contract No.\ DE-AC52-06NA25396.}

\bibliography{V-T_refs} 

\end{document}